\documentclass[graybox, secnum]{svmult}

\usepackage{amsmath}
\newcommand{\Lagr}{\mathcal{L}}

\newcommand{\Ef}{E^{\prime}}
\newcommand{\Hist}{\mathcal{H}}

\usepackage{mathptmx}       
\usepackage{helvet}         
\usepackage{courier}        
\usepackage{type1cm}        
\usepackage{makeidx}         
\usepackage{graphicx}        
\usepackage{multicol}        
\usepackage[bottom]{footmisc}
\usepackage{hyperref}        
\usepackage{soul}            
\hypersetup{colorlinks=true,urlcolor=blue}
\usepackage[square,numbers]{natbib}
\usepackage{subfigure}
\makeindex             

\begin{document}
\title*{Soft gamma-ray polarimetry with COSI using maximum likelihood analysis}
\author{John A. Tomsick\thanks{corresponding author}, Alexander Lowell, Hadar Lazar, Clio Sleator, and Andreas Zoglauer}
\authorrunning{Tomsick et al.}
\institute{John A. Tomsick \at Space Sciences Laboratory, UC Berkeley, 7 Gauss Way, Berkeley, CA 94720, USA, \email{jtomsick@berkeley.edu}
\and Alexander Lowell, Hadar Lazar, and Andreas Zoglauer \at Space Sciences Laboratory, UC Berkeley, 7 Gauss Way, Berkeley, CA 94720, USA
\and Clio Sleator \at U.S. Naval Research Laboratory, Washington, DC 20375, USA}
%
%
\maketitle
\abstract{Measurements of the linear polarization of high-energy emission from pulsars, accreting black holes, and gamma-ray bursts (GRBs) provide an opportunity for constraining the emission mechanisms and geometries (e.g., of the accretion disk, jet, magnetic field, etc.) in the sources.  For photons in the soft (MeV) gamma-ray band, Compton scattering is the most likely interaction to occur in detectors.  Compton telescopes detect multiple interactions from individual incoming photons, allowing for scattering angles to be measured.  After many photons are detected from a source, the distribution of azimuthal angles provides polarization information. While the standard method relies on binning the photons to produce and fit an azimuthal scattering angle distribution, improved polarization sensitivity is obtained by using additional information to more accurately weight each event's contribution to the likelihood statistic.  In this chapter, we describe the Compton Spectrometer and Imager (COSI) and its capabilities for polarization measurements.  We also describe the maximum likelihood technique, its application to COSI data analysis, and plans for its future use.}


\vspace{2cm}

\section{Introduction}
\label{sec:intro}

Polarization measurements provide unique diagnostics for determining emission mechanisms and source geometries, but current results give just a glimpse into the potential for what we can learn.  The small number of measurements that are available indicate that high ($>$50\%) polarization levels can occur at MeV energies, suggesting the possibility of measuring energy and time dependence of polarization to probe the physics of extreme environments near compact objects or in explosive events.  MeV sources of interest include pulsars, accreting Galactic black holes (BHs), active galactic nuclei (AGN), and gamma-ray bursts (GRBs).

For pulsars and the emission from their surrounding nebulae, a key early X-ray (2.6 and 5.2\,keV) measurement was made of the Crab nebula in the 1970s using the OSO 8 satellite.  This showed a $19.2\pm 1.0$\% level of polarization, proving a synchrotron origin for the emission \citep{novick72,weisskopf78}.  INTEGRAL and AstroSat observations of the Crab at 0.1-1\,MeV indicate even higher polarization levels of 32--98\% in soft gamma-rays \citep{dean08,forot08,moran16,vadawale18}.  The angle of polarization appears to be along the spin axis of the pulsar \citep{dean08} or to wobble around the pulsar spin axis \citep{vadawale18}, suggesting that the soft gamma-ray emission is more closely related to the pulsar than to the nebula.  

For Galactic BHs, there has been a long-standing question about the origin of the emission detected up to MeV energies \citep{grove98,mr06}. These systems often show a non-thermal component extending into the MeV bandpass.  As accreting BHs have relativistic jet outflows, the jet may be the origin of the non-thermal component \citep{mnw05}. While most of the known Galactic BHs have transient outbursts of emission, Cygnus~X-1 is persistently bright in X-rays, and its high-energy spectrum includes a non-thermal component \citep{mcconnell02}.  By combining data from long observations over several years at times when Cyg~X-1 was known to be producing jets, both instruments on INTEGRAL measured highly polarized emission ($67\pm 30$\% with IBIS and $76\pm 15$\% with SPI) above $\sim$0.4\,MeV \citep{laurent11,jourdain12}.  Meanwhile, the polarization levels below this energy are constrained to be much lower by INTEGRAL \citep{laurent11,jourdain12} and the PoGO+ balloon mission \citep{chauvin18}.  These measurements are readily explained if the emission up to $\sim$0.2\,MeV is dominated by inverse Comptonization from a quasi-spherical corona, with the geometry causing the low polarization, while the $>$0.2\,MeV emission comes from the jet and is polarized due to the collimation and structured magnetic field.  The polarization measurements provide insight into the emission geometry that is not provided by the spectral measurements alone.

MeV polarization measurements also provide unique information about AGN and can be used to probe jet composition and emission mechanisms.  For blazars, there is a long-standing question of whether the emission is of hadronic or leptonic origin. A high level of polarization ($\sim$60\%) would indicate that the emission is hadronic \citep{zb13}.  For radio galaxies (e.g., Cen~A), a high level of polarization will show that the emission is produced by a jet via synchrotron self-Compton (SSC) \citep{abdo10} rather than by inverse-Compton scattering in a hot tenuous accretion disk corona \citep{beckmann11}. The level of polarization is expected to be $\sim$5-10\% for the thermal case \citep{matt93} and up to 60\% for the SSC case \citep{krawczynski12}.  

For GRBs, polarization measurements could shed light on the engines that radiate the observed photons by comparing to theoretical predictions that show that the distributions of GRB polarizations depend on whether the emission mechanism is due to synchrotron or inverse Comptonization and also on the properties of the magnetic field in the jet \citep{toma09,gkg21}. In designing an instrument for GRB studies, it is important to consider that they occur at random times and locations on the sky, requiring an instrument with a large field of view.  With both polarization sensitivity and large field of view, Compton telescopes are well-suited for studies of GRBs.

In this chapter, we describe the use of Compton telescopes for MeV polarization measurements, emphasizing the improvements in polarization sensitivity obtained when the maximum likelihood method \citep{krawczynski11} is used rather than the standard method.  We start with an explanation of how Compton telescopes operate and considerations for polarimetry.  One important point is that the signal used to make the polarization measurement is strongest for photons that have $\sim$90$^{\circ}$ Compton scatter angles.  The previous Compton telescope, COMPTEL \citep{schoenfelder93} on the Compton Gamma-Ray Observatory (CGRO), was designed to primarily measure events with small scatter angles, and was thus not very sensitive to polarization.  The next generation of Compton telescopes are designed with 3D position resolution, allowing them to measure events with large scatter angles.  We describe a particular instrument, COSI \citep{kierans17,tomsick21}, and explain its polarization capabilities using calibration and astrophysical measurements.  In particular, a detailed polarization study was carried out for GRB~160530A using the maximum likelihood technique \citep{lowell17b,lowell17a}.  We describe the maximum likelihood technique and then end the chapter by discussing a framework for its development and use in future Compton telescope missions.

\section{Compton Telescopes and Polarization Measurements}
\label{sec:compton_telescopes}

\subsection{Operation of Compton Telescopes}
\label{sec:operation}

Compton telescopes are a powerful technology for performing imaging, spectroscopic, and polarimetric studies of photons in the 0.2-10\,MeV band.  This is due to the fact that the cross-section for Compton scattering dominates over photoabsorption and pair-production in this energy range for most detector materials of interest.  A Compton telescope works by exploiting the phenomenon of Compton scattering, whereby a photon undergoes a collision with an electron.  A canonical Compton event in a Compton telescope consists of one or more Compton scatters followed by photoelectric absorption of the scattered photon, all in the active detector volume.  

\begin{figure}[h]
\centerline{\includegraphics[height=3.3in,angle=0]{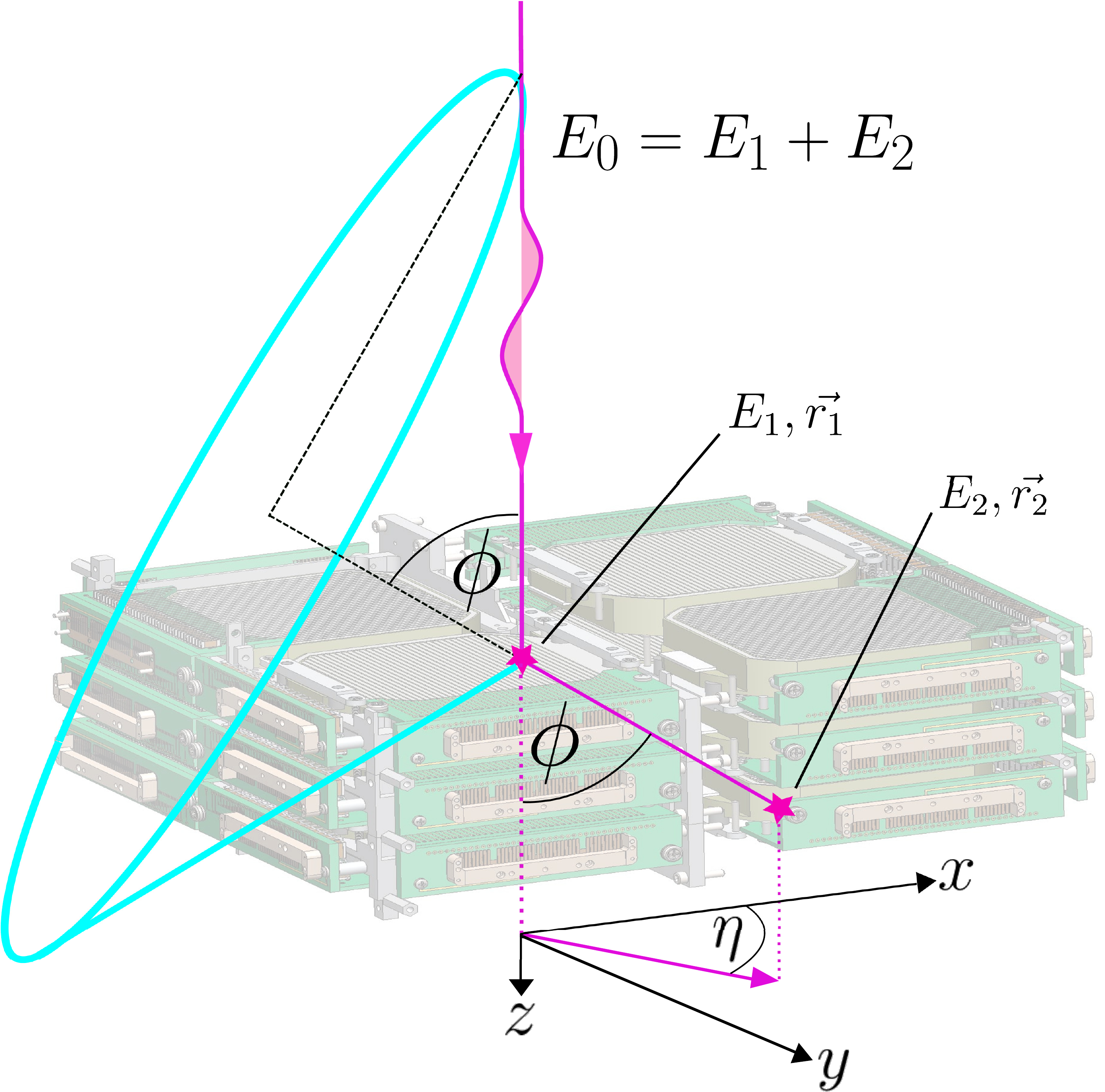}}
\vspace{0.0cm}
\caption{Illustration of a gamma-ray event with two interactions measured by detectors with 3D position measurement capabilities (the COSI instrument is shown).  Key angles for a Compton telescope are the Compton scattering angle ($\phi$) and the azimuthal scattering angle ($\eta$).  From \citep{lowell17_phd} (Figure 2.1 with one variable name change). \label{fig:axes_and_angles}}
\end{figure}

Figure~\ref{fig:axes_and_angles} provides an example of a photon traveling along the $z$ direction with initial energy $E_0$.  The photon Compton scatters with Compton scattering angle $\phi$ at $\vec{r_1}$, depositing an energy of $E_1$.  After scattering, the photon's energy is $E^{\prime} = E_0 - E_1$.  The scattered photon is then photoelectrically absorbed at $\vec{r_2}$, depositing an energy of $E_2=E^{\prime}$.  A Compton cone with half opening angle $\phi$ can now be constructed by placing the cone vertex at $\vec{r_1}$, and aligning the cone axis with the vector $\vec{r_1} - \vec{r_2}$.  The Compton cone defines a Compton circle on the sky, representing all possible points from which the photon may have originated and providing a Compton telescope's ability for imaging.  

In addition to the Compton (polar) scattering angle $\phi$, there is also an azimuthal scattering angle $\eta$ involved in the Compton scattering process.  The distribution of $\eta$ from a sample of photons originating from a beam is sensitive to the beam's polarization level and angle.  As described in Section~\ref{sec:cpol}, this forms the basis for using Compton telescopes as polarimeters in the soft gamma-ray band.

\subsection{Compton Polarimetry}
\label{sec:cpol}

\begin{figure}[h]
\centerline{\includegraphics[height=3.3in,angle=0]{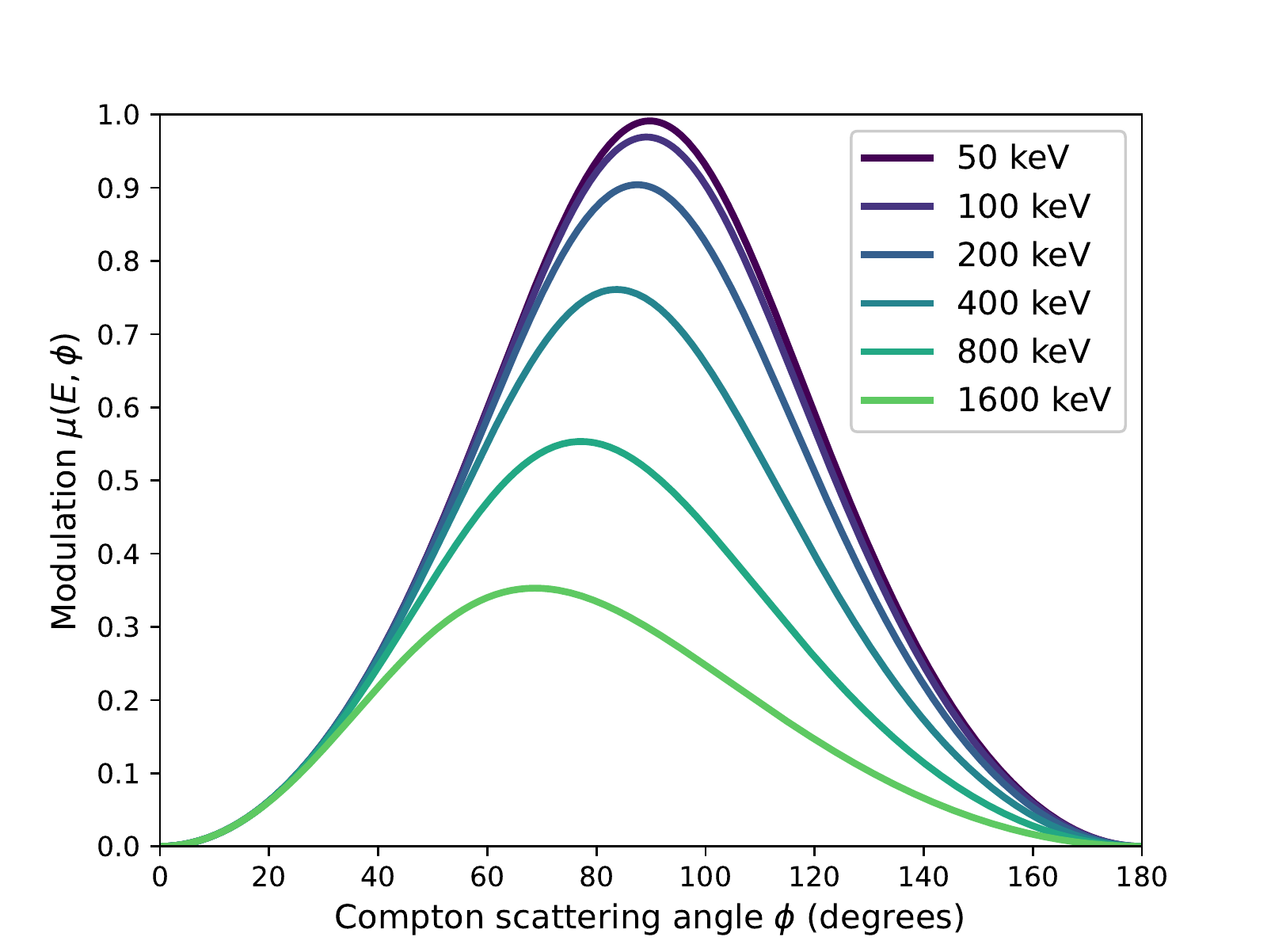}}
\vspace{0.0cm}
\caption{Modulation amplitude for six different energies as a function of Compton scattering angle. This is an instrument characteristic, and the higher the modulation amplitude, the more sensitive the instrument will be for polarization measurements. From \citep{lowell17a} (Figure 1 with one variable name change).
\label{fig:modulation}}
\end{figure}

The Klein-Nishina equation gives the differential cross-section for Compton Scattering of photons on free electrons at rest according to
\begin{equation} \label{eq:kn}
\frac{d\sigma}{d\Omega} = \frac{r_{0}^2}{2}\bigg(\frac{E^{\prime}}{E}\bigg)^2\bigg(\frac{E}{E^{\prime}} + \frac{E^{\prime}}{E} - 2\sin^2 \phi \cos^{2} \eta \bigg)~~~, 
\end{equation}
where $r_{0}$ is the classical electron radius, $E$ is the initial photon energy, $E^{\prime}$ is the scattered photon energy, $\phi$ is the Compton scattering angle, and $\eta$ is the azimuthal scattering angle defined such that $\eta = 0$ corresponds to scattering along the direction of the initial photon's electric field vector.  After some algebraic manipulation of Equation \ref{eq:kn}, the probability density function (PDF) of scattering with a particular $\eta$ takes the simple form of an offset cosine:
\begin{equation}
p(\eta;E,\phi) = \frac{1}{2\pi} \big[ 1 - \mu(E,\phi)\cos(2\eta) \big]~~~,
\label{eq:azprob}
\end{equation}
where the modulation, $\mu (E,\phi)$, is defined as
\begin{equation}
\mu (E,\phi) = \cfrac{\sin^2\phi}{\cfrac{E^{\prime}}{E} + \cfrac{E}{E^{\prime}} - \sin^2\phi}~~~.
\label{eq:mu}
\end{equation}
$E$, $E^{\prime}$, and $\phi$ are all related by the kinematic Compton scattering formula:
\begin{equation}
\Ef = \cfrac{E}{1 + \cfrac{E}{m_e c^2}\big(1 - \cos\phi\big)}~~~,
\label{eq:compton}
\end{equation}
where $m_ec^2 = 511$\,keV.  The dependence of $\mu$ on $E$ and $\phi$ is shown in Figure~\ref{fig:modulation}.  The value of $\mu$ is larger at lower energies and for Compton scattering angles near $\sim 90^{\circ}$.

When a gamma-ray beam is polarized at a level of $\Pi$, where $0 \leq \Pi \leq 1$, then a fraction $\Pi$ of the photons from the beam will have their electric field vectors aligned along a specific direction.  The other $1-\Pi$ fraction of the photons will have their electric field vectors randomly oriented.  Thus, for a photon from a beam with polarization level $\Pi$ and polarization angle $\eta_0$, Equation \ref{eq:azprob} becomes:

\begin{equation}
p(\eta;E,\phi,\Pi,\eta_0) = \frac{1}{2\pi} \big[ 1 - \Pi\mu(E,\phi)\cos(2(\eta - \eta_0)) \big]~~~.
\label{eq:idealpdf}
\end{equation}
It is clear from Equation \ref{eq:idealpdf} that photons from a polarized gamma-ray beam will preferentially scatter such that $\eta - \eta_0 = +90^{\circ}$ or $-90^{\circ}$.  

The standard method of measuring polarization with a Compton telescope involves constructing the azimuthal scattering angle distribution (ASAD), a histogram of the azimuthal scattering angles, and fitting it with a simple cosine to determine the polarization properties of the incident beam. When performing polarimetry with a real instrument, however, geometric effects such as finite position resolution and non-uniform efficiency can affect the ASAD and thus the polarization measurement. To correct for these effects, we generate a correction ASAD from a simulation of an unpolarized source and rescale it by its mean value. The source ASAD is then divided by the correction ASAD. This process corrects for geometric effects since these effects will affect the polarized and unpolarized ASADs equally.

We then fit the corrected ASAD with a simplified version of Equation \ref{eq:idealpdf}:
\begin{equation}
    P(\eta) = P_{0} + A \cos (2 (\eta- \eta_{0}))~~~,
    \label{eq:modcurve}
\end{equation}
in which $P_{0}$ is the offset, $A$ is the amplitude, and the modulation $\mu = P_{0}/A$. The polarization level $\Pi=\mu/\mu_{100}$, in which $\mu_{100}$ is the modulation of a 100\% polarized beam. Thus, constraining $A$, $P_0$, and $\eta_{0}$ allow us to infer the polarization level and angle of a polarized source.

The maximum likelihood method, described in detail in Section~\ref{sec:mlm}, improves upon the standard method because the data are not binned and additional information is used: the photon energy and Compton scatter angle implicitly weight each event's contribution to the likelihood statistic \cite{krawczynski11}. Due to this weighting, all events can be used in the analysis.

\subsection{Designing a Compton Polarimeter}
\label{sec:designing}

A good Compton polarimeter is able to measure those areas in the data space which reach the highest modulation values. According to Figure~\ref{fig:modulation}, the largest polarization modulation can be detected at low energies and large Compton scatter angles ($\sim$60--90$^{\circ}$). 

Therefore, a good Compton polarimeter needs to have a detector geometry which allows for the measurement of a wide range of Compton scatter angles covering the whole modulation peak. This criteria can be fulfilled by compact Compton telescopes (CCTs) with a detector arrangement that is approximately a single volume (such as COSI \citep{kierans17,tomsick21}) or with a scatter detector that is surrounded by an absorber detector. Instrument concepts where the absorber is close to the bottom of the tracker such as e-ASTROGAM \citep{deangelis17} and AMEGO \citep{mcenery19} are somewhat less optimized because large Compton scatter angles can only be measured at larger incidence angles. Going even further, Compton telescope designs with a large distance between the scatterer and the absorber are even less ideal to measure polarization. This is the first main reason that COMPTEL \citep{schoenfelder93}, the first successful Compton telescope in space, was not able to measure polarization, even from the bright Crab nebula and pulsar.  With two detector planes separated by $\sim$1.5\,m to enable time-of-flight background rejection, COMPTEL was unable to measure events with large Compton scattering angles.

In addition, a Compton telescope optimized for polarimetry should also allow for the measurement of Compton scatters down to low energies. This requires low-atomic number (Z) materials for the Compton scatter detectors, since for those materials Compton scattering can be the dominating interaction process down to $\sim$50\,keV. For example, the cross-overs where the Compton scattering cross-section starts to dominate of the photo effect cross-section are $\sim$57\,keV for silicon, $\sim$150\,keV for germanium, $\sim$263\,keV for cadmium zinc telluride (CZT), and $\sim$290\,keV for cesium iodide (CsI).  In addition, those low-energy interactions, which are usually spatially very nearby, need to be resolved in different voxels in the detector. This requires either very good position resolution in a larger detector or thin scatter detectors. For COMPTEL, with liquid scintillator and sodium iodide (NaI) detectors, the lower-energy threshold was $\sim$750\,keV, which is the second main reason why it was not able to measure polarization.

Furthermore, the polarization calibration is as important as the instrument design. The goal of the calibration is to understand any systematics in the azimuthal scattering angle distribution down to a level acceptable for the mission as a function of energy, incidence direction, and polarization level. The calibration also serves as the benchmark for simulations, which are needed to determine the full polarization response of the instrument later on for all possible incidence directions compared to the select few directions covered by the calibration (see Section~\ref{sec:calibration}). 

Finally, considering on-orbit observations, heavily shielded, (very) narrow-field-of-view instruments have the advantage that they can use use instrument rotations to average over systematics and on-and-off-source observations to determine the background. All-sky Compton telescopes cannot do this. Therefore, it is important to have a narrow point-spread function (PSF) in order to have ample background-only regions in the sky to understand any induced modulation in the azimuthal scattering angle by the background. This requires good angular resolution which can be achieved by detectors which have good energy and position resolution, as well as detectors which use low-to-medium-Z scatter detector materials to minimize Doppler broadening.  Doppler broadening limits the angular resolution of a Compton telescope due to scattering on a bound electron with unknown initial momentum (see \citep{zk03}).





\subsection{The Compton Spectrometer and Imager}
\label{sec:cosi}

COSI is a CCT, which operates in the 0.2--5\,MeV bandpass.  While a satellite version of COSI \citep[COSI-SMEX,][]{tomsick19,tomsick21} has been designed, we focus on the balloon-borne version of the COSI instrument.  This is because a full balloon-borne mission has been completed from integration and calibration to data collection during flight to data analysis.  In the following, we include a description of the balloon-borne version of the instrument and the polarization calibration and summarize the 2016 COSI balloon flight, focusing on the detection of GRB~160530A and the polarization study of this GRB.

\subsubsection{Instrument}
\label{sec:instrument}

\begin{figure}[h]
\centerline{\includegraphics[height=3.3in,angle=0]{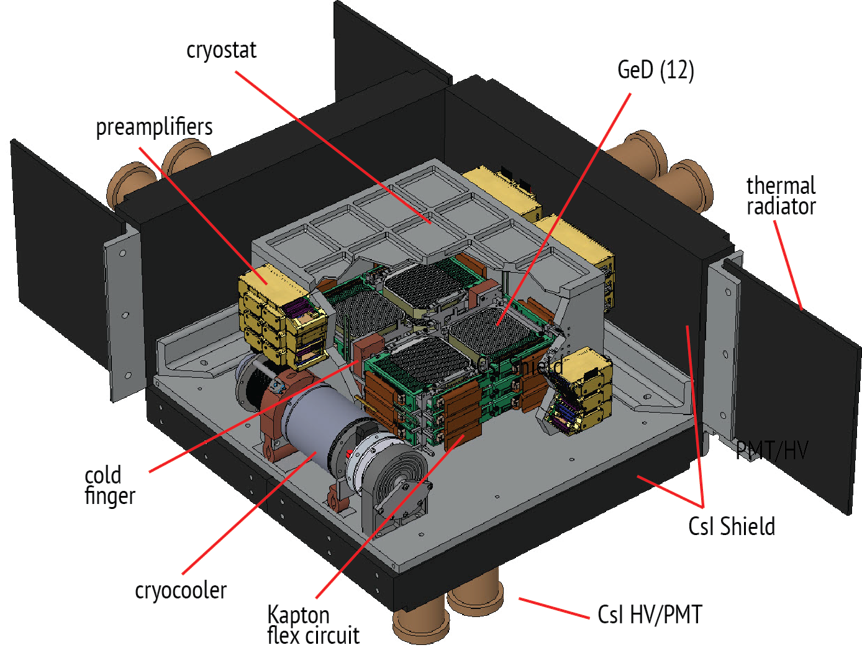}}
\vspace{-0.2cm}
\caption{A schematic of the COSI GeDs, cryostat, cryocooler, and CsI anticoincidence shields.  The schematic is for the instrument that has flown on high-altitude balloons. From \citep{kierans17} (Figure 2).
\label{fig:schematic}}
\end{figure}

The heart of COSI consists of a 2$\times$2$\times$3 array of high purity germanium double-sided strip detectors (GeDs), each with a volume of 8$\times$8$\times$1.5\,cm$^3$. The GeD array is housed in an aluminum cryostat and is cooled to cryogenic temperatures with a Sunpower Cryotel CT mechanical cryocooler, enabling ultra-long duration balloon flights since no consumables are required. The cryostat is surrounded by CsI scintillators that provide both passive and active shielding around the bottom and sides of the cryostat, as shown in Figure~\ref{fig:schematic}. The shields define COSI's field of view, which is $>$25\% of the sky.

The electrodes on each side of each GeD are segmented into 37 strips with a 2\,mm strip pitch. The strips on the anode are deposited orthogonally to those on the cathode so that the $x$-$y$ interaction position can be determined from the positions of the triggered strips. To determine the $z$ position, we use the collection time difference (CTD), or the difference in arrival times between the electrons on the anode and the holes on the cathode (see \citep{lowell16} for a description of how the CTD-depth relation is calculated). The position resolution in the $x$ and $y$ directions is equal to the strip pitch, 2\,mm, and the position resolution the $z$ direction is on average 0.2\,mm RMS \citep{lowell16}. With three-dimensional position resolution and excellent spectral resolution of 0.2\% to 1\% (depending on the photon energy), COSI is a natural polarimeter.

An ideal event consists of at least one Compton scatter and one photoabsorption, all occurring within the GeDs. Since COSI is a CCT, the volume in which the interactions take place is too small to use time of flight to determine the order of interactions. Instead, the event can be reconstructed using a variety of techniques, including Compton kinematic reconstruction \citep{bj00} and Bayesian reconstruction \cite{zoglauer05_phd}. Once the interaction order has been determined, the origin of the photon can be constrained to a circle on the sky using the classic Compton scattering formula (see Figure~\ref{fig:axes_and_angles}). At this point, events with at least two interactions in the GeDs can be used for high level analysis, including Compton polarimetry. To perform the event reconstruction and high level analysis, we use the MEGAlib software library \cite{zoglauer06}, which is specifically designed to analyze data from Compton telescopes.

Simulations are crucial for both benchmarking the instrument performance and determining the instrument response. We use the MEGAlib wrapper of Geant4 \cite{agostinelli03} to perform Monte Carlo simulations of particles and propagate them through the detailed mass model of the instrument. The simulations then go through the ``detector effects engine" \cite{sleator19}, which applies the intrinsic detector performance (e.g., finite position and energy resolution) and mimics the readout electronics (e.g., thresholds, cross-talk, dead time, etc.). At this point, the simulations closely resemble the raw data and are processed through the same pipeline: the event calibration, in which the measured parameters of pulse height, ADC, and timing are converted into the physical parameters of energy and 3D position, followed by the event reconstruction.

High level analysis of both measurements and simulations proceeds with choosing event selections, which can reduce background, increase detection sensitivity, and optimize the imaging, spectral, and polarization response. The event selections used for COSI data include the total photon energy, the initial Compton scatter angle, the distance between interactions within the detector, and the number of interaction sites within the active volume.

\subsubsection{Polarization Calibration}
\label{sec:calibration}

In order to determine COSI’s polarization response and to identify systematic deviations from an ideal sinusoidal modulation in $\eta$, it is necessary to evaluate COSI's polarization performance in the laboratory. This evaluation requires measuring a polarized beam and simulating the experimental configuration.  While producing a fully polarized beam in the laboratory is non-trivial, we are able to produce a partially polarized gamma-ray beam using a principle detailed by Lei et al. \cite{lei97}. When unpolarized photons Compton scatter, the outgoing beam is partially polarized with a polarization level given by
\begin{equation}
    \Pi = \frac{\sin^{2} \phi}{\epsilon + \epsilon^{-1} - \sin^{2}\phi}~~~,
\end{equation}
in which $\epsilon$ is the ratio of scattered photon energy to initial photon energy and $\phi$ is the Compton scattering angle. The polarization vector of the scattered beam is perpendicular to the scattering plane. We produced partially polarized gamma-ray beams in the laboratory by scattering photons from a NaI scintillator, which has an attached photomultiplier tube.  As an example, 661.7\,keV photons that scatter off the scintillator at 90$^{\circ}$ (to obtain the highest polarization level possible) produce a beam with an outgoing photon energy of 288\,keV and a polarization level of approximately $\sim$58\% \citep{lowell17_phd}. Using an active detector as a scattering surface allows us to select only events coincident between COSI and the NaI and thus reject the majority of the background. As the count rate of the scattered photons is quite low ($<$1 count\,s$^{-1}$), rejecting a large amount of the background using the coincidence technique is essential.  

\begin{figure}[ht]
\begin{subfigure}
  \centering
  \includegraphics[width=.5\linewidth]{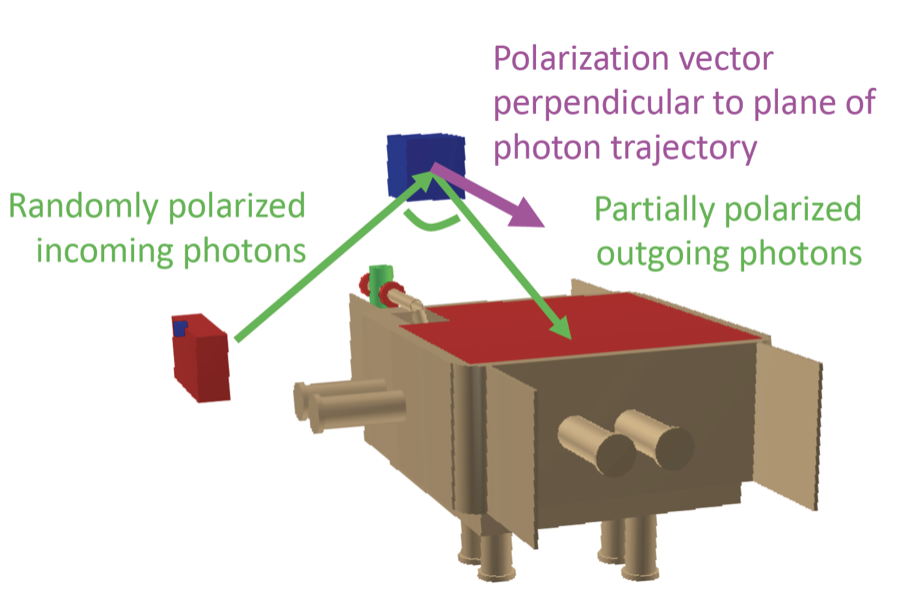}  
\end{subfigure}
\begin{subfigure}
  \centering
  \includegraphics[width=.5\linewidth]{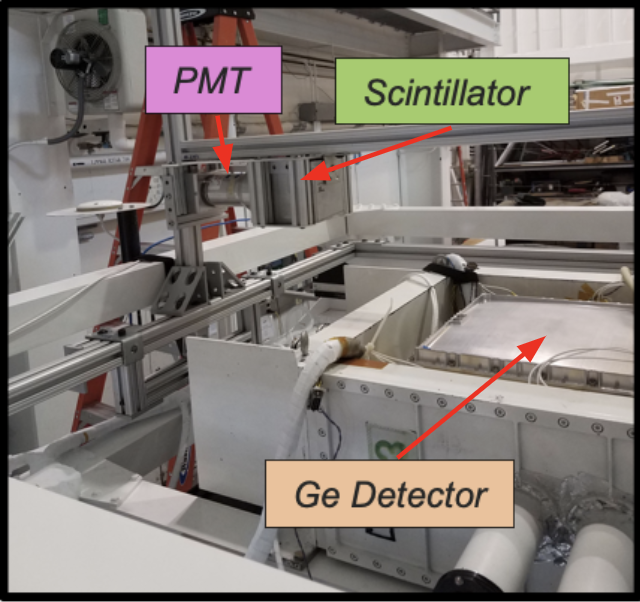}  
\end{subfigure}
\caption{(a) To produce a partially polarized beam, gamma-rays are emitted from a $^{137}$Cs source and scatter off a scintillator towards the detector. (b) The setup at Space Sciences Laboratory, Berkeley, 2019.}
\label{fig:polar}
\end{figure}

Figure~\ref{fig:polar} shows the geometrical configuration for pre-flight calibrations. We suspend the NaI scintillator above the instrument (GeDs within the cryostat surrounded by the CsI shields), while fixing a $^{137}$Cs source (661.7\,keV) at the cryostat’s level. We place a lead brick between the $^{137}$Cs source and the COSI detector system in order to prevent the direct flux from unnecessarily elevating the shield count rate. Acquiring data with the NaI scintillator in many different locations captures a range of polarization and scattering angles. 

\begin{figure}[h]
\centerline{\includegraphics[height=3.3in,angle=0]{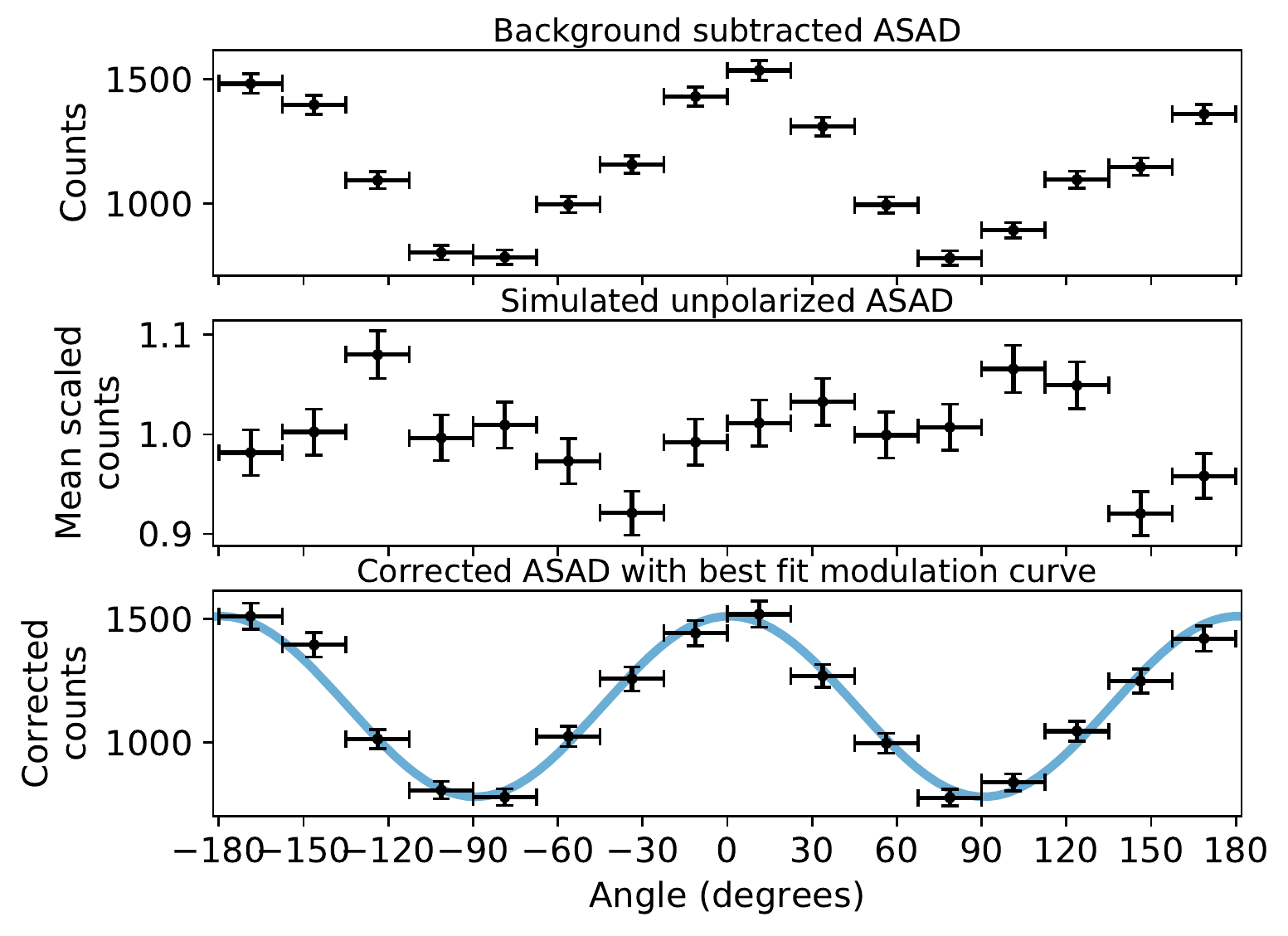}}
\caption{Azimuthal Scattering Angle Distributions (ASADs) from the COSI calibrations taken in 2016.  {\em Top:} The measured ASAD with a 295\,keV partially-polarized beam scattered from a NaI scintillator. {\em Middle:} Simulated unpolarized ASAD based on measurements with radioactive sources used to benchmark simulations.  {\em Bottom:} Corrected ASAD fitted with a sinusoid. From \citep{lowell17_phd} (Figure 5.13).
\label{fig:asad}}
\end{figure}

Azimuthal scattering angle distributions (ASADs) are generated from the calibration data.  The coincident events, which are effectively background-subtracted by calculating the number of expected chance coincidences, are used to produce azimuthal scattering angle distributions (ASADs) as shown in the top panel of Figure~\ref{fig:asad}. The ASAD is then corrected for the instrument response by being divided bin-by-bin with an ASAD of unpolarized events (Figure~\ref{fig:asad}, middle panel). The data for the unpolarized beam can be measured from direct illumination of a radioactive source at the same location as the NaI scintillator or from simulations of an unpolarized beam at the same location as NaI scintillator \citep{lowell17_phd,yang18}.

Once the fully-corrected partially-polarized ASAD is produced by dividing the measured ASAD by the scaled unpolarized ASAD (Figure~\ref{fig:asad}, bottom panel), the polarization fraction and angle can be derived by fitting the ASAD with Equation~\ref{eq:modcurve}. Figure~\ref{fig:asad} (bottom panel) shows an example of a corrected ASAD fitted with the above equation.  The calibration provides an important validation that the corrected ASAD is sinusoidal and has the correct phase ($\eta_{0}$) for the known polarization direction.  In addition, the amplitude, $A$, provides a determination of the modulation factor.  To effectively benchmark the simulated instrument response with calibrations, we can employ non-parametric statistical tests.  The Kolmogorov-Smirnov and Anderson-Darling tests are used to evaluate the hypothesis that the azimuthal scattering angle samples from the measurements and simulations were drawn from the same underlying distribution \citep{lowell17_phd}.

\subsubsection{2016 Balloon Flight and GRB 160530A}
\label{sec:flight}

On 2016 May 17, COSI was launched from Wanaka, New Zealand, on NASA's Superpressure balloon. COSI had a successful 46-day flight, circumnavigating the Earth one and a half times before landing in Peru. During the 2016 flight, COSI observed persistent astrophysical emission including the 511\,keV signature of positron annihilation from the Galactic center \cite{kierans20}\cite{siegert20} as well as the Crab, Cygnus X-1, and Centaurus A \cite{sleator19_phd}. See \cite{kierans17} for more details about the 2016 flight.

On 2016 May 30 at 07:03:46 UT, COSI detected and sent the discovery notice for the bright, long gamma-ray burst GRB~160530A \cite{tomsick16_cosi}. At the time of this observation, the COSI instrument was at geographic coordinates 56.79$^\circ$~S, 82.31$^\circ$~E and was floating at an altitude of 32.6\,km. Due to the close proximity of the South Magnetic Pole during this observation, the background was relatively high. Additionally, a relativistic electron precipitation (REP) event \cite{parks79} was occurring during the time of the GRB, introducing low-frequency variations into the background count rate. At this point in the flight, two of COSI's detectors had suffered from high voltage failures due to defects in the potting of the high voltage filters, and thus 10 out of 12 GeDs were operating at the time of the GRB.

\begin{figure}[h]
\centerline{\includegraphics[height=1.5in,angle=0]{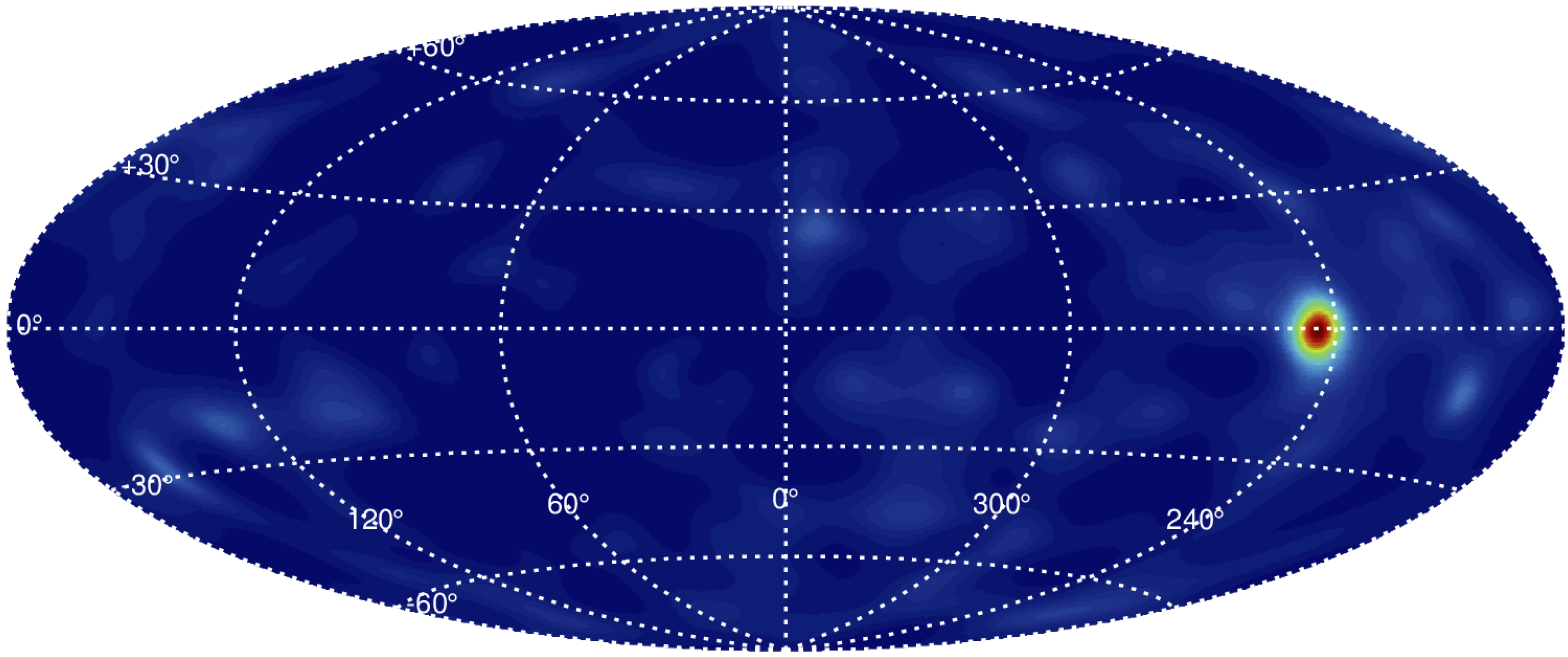}}
\caption{COSI image of GRB~160530A in the 0.1-1\,MeV energy range.  The image was produced using the list mode maximum-likelihood expectation-maximum (LM-MLEM) image deconvolution algorithm \citep{lowell17a}.  GRB~160530A is a long GRB lasting 38.9\,s that was used for a polarization study.  From \citep{sleator19_phd} (Figure 5.5).
\label{fig:grb_image}}
\end{figure}

Figure~\ref{fig:grb_image} shows the COSI image of GRB~160530A, made using 834 events and event selections that optimize imaging performance. The peak of the image is at $l=243.4^\circ$, $b=0.4^\circ$, which is the best known position of this GRB. In the local instrument coordinate system, the GRB position was 43.5$^\circ$ off-axis with an azimuth of --66.1$^\circ$. GRB 160530A was also detected by Konus-Wind and INTEGRAL/ACS \cite{svinkin16a}, both of which are constituents of the Inter-Planetary Network (IPN, \cite{hurley10}); the IPN localized this GRB to an annulus on the sky, which overlapped with the COSI position.

Spectral analysis of GRB~160530A was performed with both Konus-Wind \cite{svinkin16b} and COSI \cite{sleator19_phd}. Using 38.9\,s of Konus-Wind data between 20\,keV and 5\,MeV, fitting with a Band model resulted in $\alpha=-0.93\pm0.03$, $E_p=638^{+36}_{-33}$\,keV, and an upper limit on $\beta < -3.5$. With $E_p$ and $\beta$ fixed to the Konus-Wind best-fit values, the COSI team measured a consistent spectral shape with $\alpha=-1.14^{+0.30}_{-0.32}$, using the same 38.9\,s of data as Konus-Wind and an energy range of 100\,keV to 3\,MeV. For all simulations used in the polarization analysis, the best fit Konus-Wind Band model was given as the spectral input.  This was a bright GRB with a fluence of $(1.30\pm 0.04)\times 10^{-4}$\,erg\,cm$^{-2}$.

The COSI team measured the polarization of of GRB~160530A using both the standard method of fitting the azimuthal scattering angle distribution (ASAD) and a maximum likelihood analysis \cite{lowell17b,lowell17a}. In the standard analysis, the azimuthal scattering angles are histogrammed into the ASAD and fit with Equation~\ref{eq:modcurve}; the measured modulation $\mu$ is $\mu = A/P_0$ and the degree of polarization is $\mu/\mu_{100}$, where $\mu_{100}$ is the modulation of a 100\% polarized source. To determine the $\mu_{100}$, we simulated a 100\% polarized GRB 160530A, resulting in $\mu_{100} = 0.484\pm0.002$. We determined the event selections that optimize the minimum detectable polarization (MDP):
\begin{equation}
    {\rm MDP} = \frac{4.29}{\mu_{100}r_s}\sqrt{\frac{r_s+r_b}{t}}~~~,
    \label{eq:mdp}
\end{equation}
where $r_s$ is the source count rate, $r_b$ is the background count rate, $t$ is the observation time, and the factor of 4.29 corresponds to 99\% confidence \citep{weisskopf10}. The optimized event selections resulted in 445 total counts for the standard method analysis, 123 of which were background. Since Equation~\ref{eq:mdp} does not take systematic error into account, we generated $N=10,000$ trial data sets from an unpolarized simulation of GRB~160530A, performed the polarization analysis, and stored the resulting polarization level. The MDP is the 99th percentile of polarization levels, found to be ${\rm MDP} = 72.3\pm0.8\%$.

As described in Section \ref{sec:cpol}, we corrected for geometric effects using a correction ASAD generated from a simulation of a completely unpolarized GRB 160530A. The geometry-corrected source ASAD was then fit with Equation~\ref{eq:modcurve}, resulting in a measured polarization level of ${\rm \Pi}=33^{+33}_{-31}\%$ for GRB~160530A. This polarization level is below the MDP, indicating that COSI did not detect polarization from this GRB using the standard method.

With the maximum likelihood method and weighing events according to their contribution to the likelihood statistic, all events can be used in the analysis rather than selecting on the events that optimize the MDP in the standard method analysis. In the case of COSI's observation of GRB~160530A, 542 counts were used in the analysis, 152 of which were background. Similarly to the standard method analysis, the MDP was calculated by generating and analyzing 10,000 trial runs, yielding ${\rm MDP} = 57.5\pm0.8$\%. The geometry-corrected polarization level of the maximum likelihood analysis of GRB~160530A was ${\rm \Pi}=16^{+27}_{-16}$\%, which again was below the MDP.

In both the standard method and the maximum likelihood method, the measured polarization for GRB~160530A was below the detection limit, defined by the MDP. Thus, COSI did not detect polarized emission from GRB~160530A. We note that COSI's polarization sensitivity was reduced during this observation due to a number of factors. Two out of 12 detectors were non-operational due to a high voltage problem, resulting in a loss of 16\% of events. The REP event that occurred at the same time as the GRB and COSI's proximity to the South Magnetic Pole elevated the background count rate. Additionally, the GRB occurred 43.5$^\circ$ off-axis, leading to an effective area reduction of 22\%. Regardless, using the maximum likelihood method provides a significant improvement in polarization sensitivity. Particularly, the MDP of the maximum likelihood method is 21\% lower than that of the standard method, consistent with the case of an idealized polarimeter as reported in \cite{krawczynski11}.

\section{Maximum Likelihood Method}
\label{sec:mlm}

The standard approach to the polarization data analysis consists of generating a histogram of measured azimuthal scattering angles for qualifying events and fitting a ``modulation curve'' to the data.  While this approach is simple and effective, it disregards information that can be used to further constrain the polarization properties of the beam, such as the Compton (polar) scattering angle, and the initial photon energy.  \citet{krawczynski11} has shown that by combining the Compton scattering angle and photon energy measurements with the azimuthal scattering angle measurement in an unbinned, maximum likelihood analysis, the sensitivity of an ideal polarimeter is improved by $\sim 21\%$ over the standard approach.

The goal of the maximum likelihood method (MLM) is to find the beam polarization level $\Pi$ and angle $\eta_0$ that maximize the likelihood $\Lagr$
\begin{equation}
\Lagr = \prod_{i=1}^{N}p(\eta_i;E_i,\phi_i,\Pi,\eta_0)~~~,
\end{equation}
where $p$ is the conditional probability of measuring the azimuthal scattering angle $\eta_i$ given that we have accurately measured the energy $E_i$ and polar scattering angle $\phi_i$ of event $i$.  For event lists longer than several hundred, $\Lagr$ can easily underflow a double precision floating point number.  To mitigate this problem, the natural logarithm of the likelihood is used:
\begin{equation}
\ln \Lagr = \sum_{i=1}^{N}\ln p(\eta_i;E_i,\phi_i,\Pi,\eta_0)~~~.
\label{eq:loglikelihood}
\end{equation}
The values of $\Pi$ and $\eta_0$ that maximize $\ln \Lagr$ also maximize $\Lagr$, since the natural logarithm is a monotonically increasing function.  A hat symbol is used to denote the optimal values, i.e. $\hat{\Pi}$ and $\hat{\eta_0}$.

For an ideal polarimeter, $p(\eta_;E,\phi,\Pi,\eta_0)$ takes the simple form of Equation~\ref{eq:idealpdf}.  However, for a real polarimeter, Equation~\ref{eq:idealpdf} no longer holds due to the systematic effects of the detector system.  The complexity of the MLM thus lies in determining $p(\eta_i;E_i,\phi_i,\Pi,\eta_0)$ for each event $i$ in such a way so as to include the instrument systematics.  Here we outline a simulation based scheme for evaluating $p(\eta_i;E_i,\phi_i,\Pi,\eta_0)$:
\begin{enumerate}
\item Carry out a simulation of the instrument mass model subjected to an \textit{unpolarized} gamma-ray beam with the same coordinates and spectrum as the source under study.
\item Define a three-dimensional histogram $\Hist[E,\phi,\eta]$ indexed by energy $E$, polar scattering angle $\phi$, and azimuthal scattering angle $\eta$.  Let the number of $E$, $\phi$, and $\eta$ bins be $j$, $k$, and $l$, respectively.  This histogram will also be referred to as the ``response.''
\item For the $i^{\mathrm{th}}$ simulated event, perform the event filtering and reconstruction, determine $E_i$, $\phi_i$, and $\eta_i$, and increment the corresponding cell in $\Hist[E,\phi,\eta]$ by one.
\item The azimuthal scattering angle probability $p(\eta_i;E_i,\phi_i,\Pi,\eta_0)$ for a real event $i$ can now be computed in the following way: take a one-dimensional slice along the $\eta$ axis of $\Hist[E,\phi,\eta]$, and call this slice $g(\eta;E_{j^{\prime}},\phi_{k^{\prime}})$, where $j^{\prime}$ is the index of the $E$ bin containing $E_i$ and $k^{\prime}$ is the index of the $\phi$ bin containing $\phi_i$.  Then the conditional PDF for $\eta$ is
\begin{equation} \label{eq:newpdf}
p(\eta;E_i,\phi_i,\Pi,\eta_0) = \frac{1}{A}\big[g(\eta;E_{j^{\prime}},\phi_{k^{\prime}}) \times \frac{1}{2\pi}\big(1 + \Pi\mu(E_i,\phi_i)\cos(2(\eta - \eta_0))\big)\big],
\end{equation}
where $A$ is a normalization constant chosen so that the area under the total PDF is equal to unity.  Equation \ref{eq:newpdf} can then be evaluated at $\eta_i$ to yield $p(\eta_i;E_i,\phi_i,\Pi,\eta_0)$.
\end{enumerate}
Equation~\ref{eq:newpdf} is intuitively simple to understand; the slices $g(\eta;E,\phi)$ encode the effects pertaining to the instrument systematics, and the second term - which is just Equation~\ref{eq:idealpdf} - is the ideal PDF.  If this analysis was carried out with an ideal polarimeter, the slices  $g(\eta;E,\phi)$ would be uniform in $\eta$, and Equation~\ref{eq:newpdf} would collapse to Equation~\ref{eq:idealpdf}.  In essence, the $g(\eta;E,\phi)$ slices represent the acceptance as a function of $\eta$, and parameterized by $E$ and $\phi$.  

An alternative scheme for determining the azimuthal scattering angle probability for each event would be to perform a series of MC simulations with various $\Pi$ and $\eta_0$ and interpolate the responses during the maximization of $\ln \Lagr$ as $\Pi$ and $\eta_0$ are varied.  However, such an approach requires significantly more simulation time in order to achieve adequate statistics in every bin of each response.  Contrast this with the approach outlined above, where only a single simulation of an unpolarized source is needed, and no interpolation based on the values of $\Pi$ and $\eta_0$ is required. In our procedure, interpolation is avoided because the part of the PDF in Equation~\ref{eq:newpdf} that depends on $\Pi$ and $\eta_0$ is analytic.

\begin{figure}[h]
\centerline{\includegraphics[height=1.7in,angle=0]{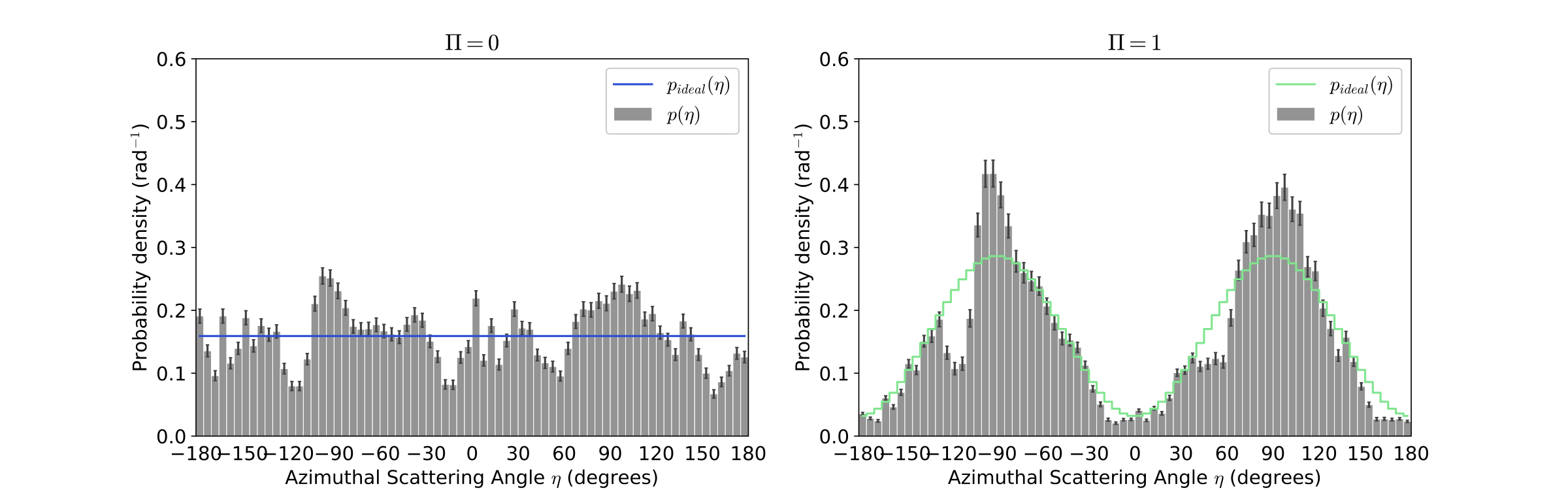}}
\caption{The full PDF used for the MLM analysis (gray bars), along with the ideal PDF (blue and green lines) for two cases $\Pi = 0$ (left) and $\Pi = 1$ (right).  The PDFs are drawn for a photon energy of 337.5\,keV and a Compton scattering angle of 92.5$^{\circ}$.  The slice $g(\eta;E,\phi)$ used for these PDFs is valid over the range $E=325-350$\,keV and $\phi = 90-95^{\circ}$.  The slice $g(\eta;E,\phi)$ used here was taken from the COSI response $\Hist[E,\phi,\eta]$ for GRB~160530A, which occurred $43.5^{\circ}$ off-axis.  Error bars are drawn on the PDFs based on the simulation statistics. From \cite{lowell17a} (Figure 2).\label{fig:pdf}}
\end{figure}

Figure~\ref{fig:pdf} shows the total PDF for the azimuthal scattering angle for a photon with $E=337.5$\,keV and $\phi=92.5^{\circ}$.  The ideal PDF is overplotted for comparison.  At this energy and Compton scattering angle, the modulation is relatively high.  On the left, where $\Pi = 0$ (unpolarized), the ideal PDF is just a constant, so the full PDF is equivalent to $g(\eta;E,\phi)$.  On the right, where $\Pi = 1$ (fully polarized), the ideal PDF is modulated, and so the full PDF is the normalized product of the modulated, ideal PDF (Equation~\ref{eq:newpdf}) with $g(\eta;E,\phi)$.  Clearly, the systematic effects of the detector system distort the PDF from its ideal shape. However, the structure of the ideal PDF still comes through in that where the ideal PDF has peaks, the probability is enhanced, and where the ideal PDF has troughs, the probability is suppressed.  Note that the response used in Figure~\ref{fig:pdf} is for the COSI observation of GRB 160530A, which occurred $43.5^{\circ}$ off-axis.

In the presence of background, the probability in Equation~\ref{eq:newpdf} must be modified to include a term that represents the background probability distribution:
\begin{equation}
p_{\textrm{total}} = f \cdot p(\eta;E,\phi,\Pi,\eta_0) + (1-f) \cdot p_{\mathrm{bkg}}(\eta;E,\phi),
\label{eq:pdftotal}
\end{equation}
where $f = (T-B)/T$ is the signal purity, $T$ is the total number of counts detected, $B$ is the estimated number of background counts in the sample, and $p_{\mathrm{bkg}}$ is the probability of measuring the azimuthal scattering angle $\eta$, given that we have accurately measured the energy $E$ and Compton scattering angle $\phi$, and that the photon originated from a source of background.  A straightforward approach for evaluating $p_{\mathrm{bkg}}$ is to generate a background response $\Hist_{\mathrm{bkg}}[E,\phi,\eta]$ with the same binning as $\Hist[E,\phi,\eta]$, filled with measured background events or simulated background events.  Each $\eta$ slice of $\Hist_{\mathrm{bkg}}[E,\phi,\eta]$ is then normalized so that the bin contents along the $\eta$ axis represent probability densities.  Finally, the background probability for event $i$ can be looked up by retrieving the contents of the bin corresponding to $\eta_{i}$, $E_i$, and $\phi_{i}$.

Once $\hat{\Pi}$ has been found, $\hat{\Pi}$ must be corrected to account for various imperfections of the detector system such as imperfect reconstruction efficiency and measurement error.  Correcting for these effects amounts to determining $\hat{\Pi}$ in the case that $\Pi = 1$, and $B = 0$.  The value of $\hat{\Pi}$ returned by the MLM algorithm under these conditions is referred to as the MLM correction factor, denoted as $\Pi_{100}$.  The corrected polarization is then given by:
\begin{equation}
\Pi = \frac{\hat{\Pi}}{\Pi_{100}}.
\label{eq:mlmpi}
\end{equation}
For an ideal polarimeter capable of perfectly reconstructing all events with perfect precision, $\Pi_{100} = 1$.  In reality, some events will be improperly reconstructed and yield a random value for $\eta$, which effectively reduces the measured polarization level.  Additionally, the measurement error on the azimuthal scattering angle will also reduce the measured polarization level.

One approach is to use the MINUIT minimizer \citep{jr75} to determine $\hat{\Pi}$ and $\hat{\eta_0}$, and MINOS (a MINUIT routine) to determine the errors for these parameters.  This uses confidence contours in the 2D $\Pi$-$\eta_0$ space along paths of constant $2\Delta \log \Lagr$, where $2\Delta \log \Lagr$ is twice the difference between the maximum log likelihood and the log likelihood of a trial point.  This quantity is asymptotically distributed as $\chi^2$ \citep{wilks38}, so the confidence level corresponding to a particular value of $2\Delta \log \Lagr$ can be calculated using a $\chi^2$ distribution with two degrees of freedom.

However, the signal purity $f$ has an associated uncertainty stemming from the Poisson distributions underlying $T$ and $B$, which will create additional uncertainty on $\Pi$.  The MINOS errors do not reflect this source of error, because $f$ is held constant during the minimization. Moreover, it is also possible that $\Pi_{100}$ will have a non-negligible uncertainty.  To determine the total uncertainty on the measured, corrected polarization level $\Pi$, the probability distribution of $\Pi$ can be approximated by repeatedly simulating the observation.  For each simulated observation, the event list is bootstrap resampled and a value of $f$ is drawn from its associated probability distribution.  Then, the minimizer is run to determine $\hat{\Pi}$ and $\hat{\eta_0}$, and $\hat{\Pi}$ is divided by a value of $\Pi_{100}$ drawn from its associated probability distribution.  The resulting distribution of $\Pi$ from the simulated observations can then be analyzed numerically to determine confidence intervals or upper limits.

The MLM has two main advantages over the SM.  First, more information is used per event.  In the SM, only the azimuthal scattering angles of qualifying events are considered.  In the MLM, the photon energy $E$ and Compton scattering angle $\phi$ are considered as well.  Effectively, each event's contribution to the likelihood statistic is implicitly weighted by Equation~\ref{eq:mu} (Figure~\ref{fig:modulation}), which is a function of $E$ and $\phi$.  Second, for a realistic observation, the MLM can use more counts in the analysis.  Consider that the first step in the SM is to optimize the statistical MDP (Equation~\ref{eq:mdp}).  This amounts to choosing event selections on $E$ and $\phi$ which yield as high a value of $\mu_{100}$ as possible, while still accepting (rejecting) as many source (background) counts as possible\footnote{$E$ and $\phi$ are the most meaningful selections in this context, but other event parameters can and should be optimized as well.}.  In the MLM however, events with any energy or Compton scattering angle can be used.  Therefore, events that were removed during the optimization of the SM analysis can now be included.  Although these events generally represent points on the $\mu$ profile (Equation~\ref{eq:mu}, Figure~\ref{fig:modulation}) corresponding to lower modulation, they are still meaningful contributors to the likelihood statistic.

Both the SM and MLM only consider the scattering parameters of the first Compton scatter.  The MLM can be extended in order to extract additional polarization information from subsequent Compton scatters, which occur frequently in compact Compton telescopes.  The method employs a matrix transformation that describes the change in the Stokes vector as a result of Compton scattering.  After each Compton scatter, the Stokes vector is transformed, a new probability distribution is computed for $\eta$, and the MLM approach is applied to the subsequent Compton scatter.  Therefore, for a sample of $N$ photons, more than $N$ Compton scatterings can be used to constrain the polarization parameters of the incoming beam, thus improving the polarization sensitivity beyond the MLM.  This method has been explored using Monte Carlo simulations of an idealized detector \citep{lowell17_phd}, but not yet with a real detector system.  

\section{Framework for Polarization Measurements for Next-Generation Compton Telescopes}
\label{sec:framework}

As discussed in Section~\ref{sec:mlm}, \citep{lowell17b} and \citep{lowell17a} demonstrate that when the MLM is chosen over SM, the MDP improves by $\sim$21\%. However, these methods may have a spectral dependence and may vary over time. This motivates time-resolved simultaneous spectral and polarization measurements, such as those conducted by Fermi-GBM and POLAR, a dedicated polarimeter launched in 2016 to the Chinese space laboratory Tiangong-2 (TG-2) \citep{burgess19}. Simultaneous fitting automatically accounts spectral uncertainties in polarization results. More next-generation Compton instruments will benefit from inferring simultaneous spectral and polarization measurements, and here we outline how such measurements can be incorporated into the MLM.  Transient sources allow for some simplifications, and we deal with transient and persistent sources separately.

\subsection{Transient Sources}
\label{sec:transients}

\begin{figure}[h]
\centerline{\includegraphics[height=3.5in,angle=0]{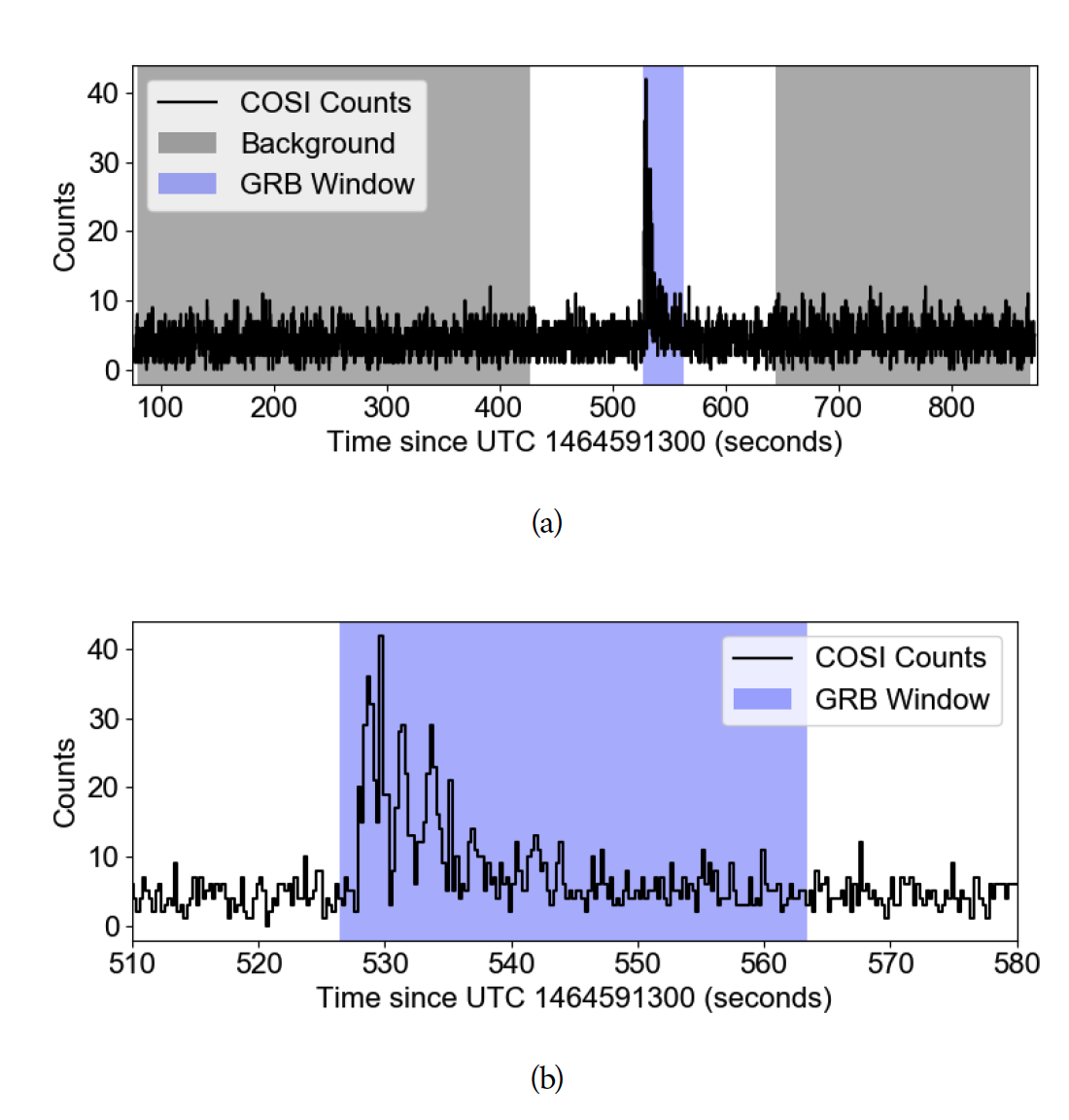}}
\vspace{-0.2cm}
\caption{(a) The COSI event light curve for GRB~160530A, illustrating the use of the data from the times before and after the GRB window for background subtraction. (b) Zoom-in of the light curve showing the GRB. From \citep{sleator19_phd} (Figure 5.6).}
\label{fig:light_curve}
\end{figure}

To outline the methodology, it is first necessary to construct likelihood functions for both spectral and polarization measurements. As both the spectra and ASADs are generated with detector counts, the source data are Poisson-distributed. The fact that it is possible to obtain direct measurements of the background before and after a burst or flare from a transient source can greatly simplify the analysis for transient sources. In Figure~\ref{fig:light_curve}, the GRB~160530A light curve that includes the events selected for spectroscopy is shown, illustrating the off-source times.  For a polarization study, a similar count rate history would be made from events selected for polarization. To select the off-source times, for every count rate history, it is necessary to determine the minimum time scale above the Poisson noise floor in which variability is introduced in the data. This can be automated via Bayesian blocks. An estimate of background counts $B_{i}$ is then given by fitting a third-order polynomial to the off-source regions \citep{lowell17b}. The uncertainties ($\sigma_{Bi}$) are determined from standard Gaussian uncertainty propagation.

For each spectral and each polarization measurement collected by COSI, the total count data in the $i^{\rm th}$ bin ($N_{i}$) is a mixture of the Poisson-distributed source ($s_{i}$) and Gaussian-distributed background ($b_{i}$) events. Thus, a probability distribution function can be assigned to each bin given $N_{i}$, $B_{i}$, and $\sigma_{Bi}$. By multiplying a Poisson distribution for data counts and a Gaussian distribution for background counts, the probability in each bin can be modeled as
\begin{equation}
    p_{PG} (N_{i} | s_{i}, b_{i}, B_{i}, \sigma _{Bi}) = p_{P} (N_{i} | s_{i} + b_{i}) p_{G} (b_{i}, \sigma _{Bi})~~~,
\end{equation}
with a total likelihood of the observation being 
\begin{equation}
    \Lagr = \Pi _{i=1} ^{N_{\rm bin}} p_{PG} (N_{i} | s_{i}, B_{i})~~~.
\end{equation}
Both polarization and spectral likelihoods take on these Poisson-Gaussian forms, with the full joint likelihood being a product of the two likelihoods. The polarization is thus inferred by maximizing this joint likelihood. The end-product is therefore the values for the spectral parameters (e.g., energy of the peak and flux of the Band function in the case of GRBs), and polarization parameters for a given Band function (polarization degree and angle). 

\subsection{Persistent Sources}
\label{sec:persistent}

The treatment of persistent signals needs to account for the reality that off-source and on-source information are not separate in time. Compton telescopes such as COSI record individual triggers in the position sensitive active detector volume, which are then used to perform event reconstruction by considering the deposited energy and the kinematics of Compton scattering. These recorded measurements store the parameters of the total photon energy deposited, Compton scattering angle $\phi \in [0, 180 ^{\circ}]$ , polar scattering angle $\psi \in [0, 180^{\circ}]$, azimuthal scattering angle $\chi \in [-180^{\circ}, 180^{\circ}]$, and a time tag. In order to model the number of counts in a data space bin \{$\phi \psi \chi t$\}, there are two approaches that can be implemented: (i) model-fitting and (ii) Richardson-Lucy deconvolution. COSI has conducted extensive analyses utilizing both approaches for the 511\,keV positron annihilation sky \citep{siegert20}. This section describes the general approach, including the amendments that could be made to include polarization. 

 The model $m_{\{\phi \psi \chi \}}$ is a predicted count rate in the Compton data space (CDS, \cite{schoenfelder93}), which is a combination of a sky model and a background model. As this is a counting experiment, the likelihood is pure Poisson. Without inputting polarization signatures, this sky model is linear, such that
\begin{equation}
    m_{\phi \psi \chi}= \alpha * m_{\phi \psi \chi}^{sky} + \beta * m_{\phi \psi \chi}^{bg}~~~,
    \label{eq:linear_model}
\end{equation}
where $\alpha$ and $\beta$ are the source and background scaling parameters \citep{siegert20}.  Defining $d$ to be the measured counts for each data space bin \{$\phi \psi \chi $\}, the likelihood is thus
\begin{equation}
    \Lagr (d| m(\alpha, \beta))= \Pi _{\phi \psi \chi} \frac{m^{d} e^{-m}}{d!}~~~.
    \label{eq:persistent_lik}
\end{equation}

As polarization is determined by the modulation of the ASAD, considering the polarization measurements in such analysis means that the azimuthal scattering angle ($\chi$) is now variable in the CDS. The effect is that Equation~\ref{eq:linear_model} is no longer expressed as a linear function because it no longer simply varies by the amplitude $\alpha$. Instead, each sky model is a function of polarization level $P$ and angle $A$, such that $\alpha m_{\phi \psi \chi }^{sky} \rightarrow \alpha(P,A) m_{\phi \psi \chi }^{sky} (P,A)$. However, this remains a counting experiment, and thus the Poission-distributed likelihood equation (Equation~\ref{eq:persistent_lik}) remains the same. The model counts vector is what changes with the addition of polarization, such that: 
\begin{equation}
    \Lagr (d| m(\alpha, \beta, P,A))= \Pi _{\phi \psi \chi} \frac{[m (P,A)]^{d} e^{-[m (P,A)]}}{d!}~~~.
    \label{eq:persistent_lik2}
\end{equation}

To infer spectral and polarization measurements simultaneously means extending the CDS further to a Compton data space with energy (CDSE). We will then have a model $m_{\{\phi \psi \chi E\}}$. An energy redistribution matrix file (RMF) is required to convert the spectral shape to the CDSE. The sky model changes such that $m_{\{\phi \psi \chi E\}}^{sky} = R * m_{\{\phi \psi \chi E\}}(P,A,p)$, where $R$ is the RMF, and $p$ is a set of spectral parameters, i.e., the centroid or width of the distribution. The form of the likelihood remains the same, with once more the model counts changing:
 \begin{equation}
    \Lagr (d| m(\alpha, \beta, P, A, p))= \Pi _{\phi \psi \chi E} \frac{[m (P,A,p)]^{d} e^{-[m (P,A,p)]}}{d!}~~~.
    \label{eq:persistent_lik3}
\end{equation}

Finally, we want to include timing or pointing information because the RMF changes depending on the aspect angle of the instrument with respect to the source. Especially for balloon instruments, the background might also not be constant as a function of time. This requires the CDS to be extended once more to be CDSET, such that the model counts are now $m_{\{\phi \psi \chi E t\}}$, with
\begin{equation}
    L (d| m(\alpha, \beta, P, A, p, t))= \Pi _{\phi \psi \chi E t} \frac{[m (P,A,p,t)]^{d} e^{-[m (P,A,p,t)]}}{d!}~~~.
    \label{eq:persistent_lik4}
\end{equation}
The total model is determined by maximizing this Poisson likelihood. These model counts are the events used to produce energy spectra and ASADs, with $d$ being the number of received photons in the signal region per bin. 

\section{Conclusions}
\label{fig:conclusions}

In this chapter, we have described instrumentation and data analysis relevant to making polarization measurements in the soft gamma-ray band.  Optimizing the instrumentation for polarization means a design that allows for measurements of Compton scattering angles near $90^{\circ}$ as well as the ability to detect Compton scatter events down to low energies.  As an example, the COSI instrument uses 3D detectors that use a relatively low-Z material (germanium) to measure low-energy interactions. On the topic of data analysis, we review work to implement the MLM method for polarization measurements of GRB 160530A, which provides a significant improvement over the standard method.  In the future, we expect further improvements to be possible.  One extension of the MLM method is to consider Compton scattering interactions beyond the first interaction to extract additional information (end of Section~\ref{sec:mlm}).  In addition, we describe a framework for polarization measurement that incorporates spectral information into the maximum likelihood framework.

\section{Acknowledgements}

The authors would like to thank Dr. Thomas Siegert for the stimulating conversations that shaped the framework for future polarization measurements of persistent sources.  We also thank Prof. Steven Boggs for suggestions that led to the improvement of this chapter.  COSI is supported through NASA APRA grant 80NSSC19K1389.


\begin{thebibliography}{}

\bibitem[\protect\astroncite{{Abdo} et~al.}{2010}]{abdo10}
{Abdo}, A.~A., {Ackermann}, M., {Ajello}, M., et~al.\  2010, ApJ, 719, 1433

\bibitem[\protect\astroncite{Agostinelli et~al.}{2003}]{agostinelli03}
Agostinelli, S., Allison, J., Amako, K., et~al.\  2003, Nuclear Instruments and
  Methods in Physics Research Section A: Accelerators, Spectrometers, Detectors
  and Associated Equipment, 506, 250

\bibitem[\protect\astroncite{{Beckmann} et~al.}{2011}]{beckmann11}
{Beckmann}, V., {Jean}, P., {Lubi{\'n}ski}, P., {Soldi}, S., \& {Terrier}, R.
  2011, A\&A, 531, A70

\bibitem[\protect\astroncite{{Boggs} \& {Jean}}{2000}]{bj00}
{Boggs}, S.~E., \& {Jean}, P.  2000, A\&AS, 145, 311

\bibitem[\protect\astroncite{{Burgess} et~al.}{2019}]{burgess19}
{Burgess}, J.~M., {Kole}, M., {Berlato}, F., et~al.\  2019, A\&A, 627, A105

\bibitem[\protect\astroncite{{Chauvin} et~al.}{2018}]{chauvin18}
{Chauvin}, M., {Flor{\'e}n}, H.~G., {Friis}, M., et~al.\  2018, Nature
  Astronomy, 2, 652

\bibitem[\protect\astroncite{{De Angelis} et~al.}{2017}]{deangelis17}
{De Angelis}, A., {Tatischeff}, V., {Tavani}, M., et~al.\  2017, Experimental
  Astronomy, 44, 25

\bibitem[\protect\astroncite{{Dean} et~al.}{2008}]{dean08}
{Dean}, A.~J., {Clark}, D.~J., {Stephen}, J.~B., et~al.\  2008, Science, 321,
  1183

\bibitem[\protect\astroncite{{Forot} et~al.}{2008}]{forot08}
{Forot}, M., {Laurent}, P., {Grenier}, I.~A., {Gouiff{\`e}s}, C., \& {Lebrun},
  F.  2008, ApJ, 688, L29

\bibitem[\protect\astroncite{{Gill} et~al.}{2021}]{gkg21}
{Gill}, R., {Kole}, M., \& {Granot}, J.  2021, arXiv e-prints,
  arXiv:2109.03286

\bibitem[\protect\astroncite{{Grove} et~al.}{1998}]{grove98}
{Grove}, J.~E., {Johnson}, W.~N., {Kroeger}, R.~A., et~al.\  1998, ApJ, 500,
  899

\bibitem[\protect\astroncite{{Hurley} et~al.}{2010}]{hurley10}
{Hurley}, K., {Golenetskii}, S., {Aptekar}, R., et~al.\  2010,
\newblock in Deciphering the Ancient Universe with Gamma-ray Bursts, ed. N.
  {Kawai}, S. {Nagataki}, Vol. 1279, American Institute of Physics Conference
  Series, 330

\bibitem[\protect\astroncite{{James} \& {Roos}}{1975}]{jr75}
{James}, F., \& {Roos}, M.  1975, Computer Physics Communications, 10, 343

\bibitem[\protect\astroncite{{Jourdain} et~al.}{2012}]{jourdain12}
{Jourdain}, E., {Roques}, J.~P., {Chauvin}, M., \& {Clark}, D.~J.  2012, ApJ,
  761, 27

\bibitem[\protect\astroncite{{Kierans} et~al.}{2017}]{kierans17}
{Kierans}, C.~A., {Boggs}, S.~E., {Chiu}, J.-L., et~al.\  2017, INTEGRAL
  Workshop Proc., arXiv:1701.05558

\bibitem[\protect\astroncite{{Kierans} et~al.}{2020}]{kierans20}
{Kierans}, C.~A., {Boggs}, S.~E., {Zoglauer}, A., et~al.\  2020, ApJ, 895, 44

\bibitem[\protect\astroncite{{Krawczynski}}{2011}]{krawczynski11}
{Krawczynski}, H.,  2011, Astroparticle Physics, 34, 784

\bibitem[\protect\astroncite{{Krawczynski}}{2012}]{krawczynski12}
{Krawczynski}, H.,  2012, ApJ, 744, 30

\bibitem[\protect\astroncite{{Laurent} et~al.}{2011}]{laurent11}
{Laurent}, P., {Rodriguez}, J., {Wilms}, J., et~al.\  2011, Science, 332, 438

\bibitem[\protect\astroncite{{Lei} et~al.}{1997}]{lei97}
{Lei}, F., {Dean}, A.~J., \& {Hills}, G.~L.  1997, Space Science Reviews, 82,
  309

\bibitem[\protect\astroncite{{Lowell}}{2017}]{lowell17_phd}
{Lowell}, A.,  2017,
\newblock Ph.D. thesis, University of California, Berkeley

\bibitem[\protect\astroncite{{Lowell} et~al.}{2016}]{lowell16}
{Lowell}, A.~W., {Boggs}, S., {Chiu}, J.~L., et~al.\  2016, SPIE Proceedings,
  9915, 99152H

\bibitem[\protect\astroncite{{Lowell} et~al.}{2017a}]{lowell17a}
{Lowell}, A.~W., {Boggs}, S.~E., {Chiu}, C.~L., et~al.\  2017a, ApJ, 848, 120

\bibitem[\protect\astroncite{{Lowell} et~al.}{2017b}]{lowell17b}
{Lowell}, A.~W., {Boggs}, S.~E., {Chiu}, C.~L., et~al.\  2017b, ApJ, 848, 119

\bibitem[\protect\astroncite{{Markoff} et~al.}{2005}]{mnw05}
{Markoff}, S., {Nowak}, M.~A., \& {Wilms}, J.  2005, ApJ, 635, 1203

\bibitem[\protect\astroncite{{Matt}}{1993}]{matt93}
{Matt}, G.,  1993, MNRAS, 260, 663

\bibitem[\protect\astroncite{{McClintock} \& {Remillard}}{2006}]{mr06}
{McClintock}, J.~E., \& {Remillard}, R.~A.  2006,
\newblock {Black hole binaries},
\newblock  157--213

\bibitem[\protect\astroncite{{McConnell} et~al.}{2002}]{mcconnell02}
{McConnell}, M.~L., {Zdziarski}, A.~A., {Bennett}, K., et~al.\  2002, ApJ, 572,
  984

\bibitem[\protect\astroncite{{McEnery} et~al.}{2019}]{mcenery19}
{McEnery}, J., {van der Horst}, A., {Dominguez}, A., et~al.\  2019,
\newblock in BAAS, Vol.~51,  245

\bibitem[\protect\astroncite{{Moran} et~al.}{2016}]{moran16}
{Moran}, P., {Kyne}, G., {Gouiff{\`e}s}, C., et~al.\  2016, MNRAS, 456, 2974

\bibitem[\protect\astroncite{{Novick} et~al.}{1972}]{novick72}
{Novick}, R., {Weisskopf}, M.~C., {Berthelsdorf}, R., {Linke}, R., \& {Wolff},
  R.~S.  1972, ApJ, 174, L1

\bibitem[\protect\astroncite{{Parks} et~al.}{1979}]{parks79}
{Parks}, G.~K., {Gurgiolo}, C., \& {West}, R. , 1979, {Geophysical Research Letters},
  6, 393

\bibitem[\protect\astroncite{{Schoenfelder} et~al.}{1993}]{schoenfelder93}
{Schoenfelder}, V., {Aarts}, H., {Bennett}, K., et~al.\  1993, ApJS, 86, 657

\bibitem[\protect\astroncite{{Siegert} et~al.}{2020}]{siegert20}
{Siegert}, T., {Boggs}, S.~E., {Tomsick}, J.~A., et~al.\ 2020, ApJ, 897, 45

\bibitem[\protect\astroncite{{Sleator}}{2019}]{sleator19_phd}
{Sleator}, C.,  2019,
\newblock Ph.D. thesis, University of California, Berkeley

\bibitem[\protect\astroncite{{Sleator} et~al.}{2019}]{sleator19}
{Sleator}, C.~C., {Zoglauer}, A., {Lowell}, A.~W., et~al.\  2019, Nuclear
  Instruments and Methods in Physics Research A, 946, 162643

\bibitem[\protect\astroncite{{Svinkin} et~al.}{2016a}]{svinkin16a}
{Svinkin}, D., {Golenetskii}, S., {Aptekar}, R., et~al.\  2016a, GRB
  Coordinates Network, 19476, 1

\bibitem[\protect\astroncite{{Svinkin} et~al.}{2016b}]{svinkin16b}
{Svinkin}, D., {Golenetskii}, S., {Aptekar}, R., et~al.\  2016b, GRB
  Coordinates Network, 19477, 1

\bibitem[\protect\astroncite{{Toma} et~al.}{2009}]{toma09}
{Toma}, K., {Sakamoto}, T., {Zhang}, B., et~al.\  2009, ApJ, 698, 1042

\bibitem[\protect\astroncite{{Tomsick} et~al.}{2019}]{tomsick19}
{Tomsick}, J., {Zoglauer}, A., {Sleator}, C., et~al.\  2019,
\newblock in BAAS, Vol.~51, ~98

\bibitem[\protect\astroncite{{Tomsick}}{2016}]{tomsick16_cosi}
{Tomsick}, J.~A.,  2016, GRB Coordinates Network, 19473, 1

\bibitem[\protect\astroncite{{Tomsick} et~al.}{2021}]{tomsick21}
{Tomsick}, J.~A., {Boggs}, S.~E., {Zoglauer}, A., et~al.\  2021, arXiv
  e-prints,  arXiv:2109.10403

\bibitem[\protect\astroncite{{Vadawale} et~al.}{2018}]{vadawale18}
{Vadawale}, S.~V., {Chattopadhyay}, T., {Mithun}, N.~P.~S., et~al.\  2018,
  Nature Astronomy, 2, 50

\bibitem[\protect\astroncite{{Weisskopf} et~al.}{2010}]{weisskopf10}
{Weisskopf}, M.~C., {Elsner}, R.~F., \& {O'Dell}, S.~L.  2010,
\newblock in Space Telescopes and Instrumentation 2010: Ultraviolet to Gamma
  Ray, ed. M. {Arnaud}, S.~S. {Murray}, T. {Takahashi}, Vol. 7732, Society of
  Photo-Optical Instrumentation Engineers (SPIE) Conference Series,  77320E

\bibitem[\protect\astroncite{{Weisskopf} et~al.}{1978}]{weisskopf78}
{Weisskopf}, M.~C., {Silver}, E.~H., {Kestenbaum}, H.~L., {Long}, K.~S., \&
  {Novick}, R.  1978, ApJ, 220, L117

\bibitem[\protect\astroncite{Wilks}{1938}]{wilks38}
Wilks, S.~S.,  1938, The Annals of Mathematical Statistics, 9, 60

\bibitem[\protect\astroncite{{Yang} et~al.}{2018}]{yang18}
{Yang}, C.~Y., {Lowell}, A., {Zoglauer}, A., et~al.\  2018,
\newblock in Space Telescopes and Instrumentation 2018: Ultraviolet to Gamma
  Ray, ed. J.-W.~A. {den Herder}, S. {Nikzad}, K. {Nakazawa}, Vol. 10699,
  Society of Photo-Optical Instrumentation Engineers (SPIE) Conference Series,
  106992K

\bibitem[\protect\astroncite{{Zhang} \& {B{\"o}ttcher}}{2013}]{zb13}
{Zhang}, H., \& {B{\"o}ttcher}, M.  2013, ApJ, 774, 18

\bibitem[\protect\astroncite{{Zoglauer}}{2019}]{zoglauer05_phd}
{Zoglauer}, A.,  2019,
\newblock Ph.D. thesis, Technische

\bibitem[\protect\astroncite{{Zoglauer} et~al.}{2006}]{zoglauer06}
{Zoglauer}, A., {Andritschke}, R., \& {Schopper}, F.  2006, New Astronomy
  Reviews, 50, 629

\bibitem[\protect\astroncite{{Zoglauer} \& {Kanbach}}{2003}]{zk03}
{Zoglauer}, A., \& {Kanbach}, G.  2003,
\newblock in X-Ray and Gamma-Ray Telescopes and Instruments for Astronomy., ed.
  J.~E. {Truemper}, H.~D. {Tananbaum}, Vol. 4851, SPIE Proceedings, 1302

\end{thebibliography}
\end{document}